\documentclass[letterpaper,english,aps,nofootinbib,showkeys]{revtex4-1}
\usepackage[T1]{fontenc}
\usepackage[utf8]{inputenc}
\usepackage{amsmath}
\usepackage{amssymb}
\usepackage{color}

\makeatletter

\pdfpageheight\paperheight
\pdfpagewidth\paperwidth

\providecommand{\tabularnewline}{\\}

\makeatother

\usepackage{babel}
\begin{document}

\preprint{}

\title{A topological charge of black holes}

\author{Yu Tian}
\email{ytian@ucas.ac.cn}

\affiliation{School of Physics, University of Chinese Academy of Sciences, Beijing
100049, China}

\affiliation{Center for Gravitation and Cosmology, College of Physical Science
and Technology, Yangzhou University, Yangzhou 225009, China}

\affiliation{Institute of Theoretical Physics, Chinese Academy of Sciences, Beijing
100190, China}

\date{\today}
\begin{abstract}
The topological charge of a maximally symmetric black hole naturally
arises in holography, which can be viewed as the last charge
of the black hole in the sense that it together with all other known
charges satisfies the holographic Gibbs-Duhem-like relation as a completeness
relation. It is then observed that this topological charge, which we have missed before, fits perfectly
into the black hole thermodynamics from two different approaches, not
only in Einstein's gravity, but also in the general Lovelock-Maxwell
theory.
\end{abstract}

\maketitle

\section{Introduction}

Black hole physics is one of the everlasting and most interesting
topics in gravitation and related fields. On one hand, black holes
are so simple objects that stationary black holes can be described
by a small number of macroscopic parameters, which we have learned from
various forms of no-hair theorems, and based on which we have black
hole thermodynamics. On the other hand, black holes are so complicated
that there are exceedingly rich phenomena and problems related to
them, such as Hawking radiation and information loss paradox, Bekenstein-Hawking
entropy and its (possible) statistical origin, Penrose process and
super-radiance, black hole perturbations and quasi-normal modes, coalescence
of binary black holes, hairy black holes, black hole phase transitions,
black holes in AdS/CFT and holography, etc.

In this paper, we focus on the simple face of black holes. For specification
and briefness, we only investigate in detail the maximally symmetric
black holes in the Einstein-Maxwell theory, leaving more general cases
in the Discussion. As is well known, the maximally symmetric black
hole solutions in the Einstein-Maxwell theory are
\begin{eqnarray}
ds_{d+1}^{2} & = & \frac{dr^{2}}{f(r)}-f(r)dt^{2}+r^{2}d\Omega_{d-1}^{(k)2},\nonumber \\
f(r) & = & k+\frac{r^{2}}{\ell^{2}}-\frac{2M}{r^{d-2}}+\frac{Q^{2}}{r^{2d-4}},\nonumber \\
d\Omega_{d-1}^{(k)2} & = & \hat{g}_{ij}^{(k)}(x)dx^{i}dx^{j},\nonumber \\
A & \sim & -\frac{Q}{r^{d-2}}dt,\label{eq:metric}
\end{eqnarray}
where $\ell$ is the AdS radius (related to the cosmological constant)
{and $k$ stands for the spatial curvature of the black hole. Specifically, $k>0$, $k=0$ and $k<0$ give a spherical, planar and hyperbolic geometry, respectively, and one usually takes $k=1,0,-1$ for convenience.}
Here and hereafter we use ``$\sim$'' to denote equality up to some
unimportant constant factor. It follows the first law of black hole
thermodynamics
\begin{equation}
d\tilde{M}=T_{H}dS+\Phi dQ
\end{equation}
with
\begin{eqnarray}
S & = & \frac{\Omega_{d-1}^{(k)}r_{h}^{d-1}}{4},\label{entropy} \\
T_{H} & = & \frac{f_{h}^{\prime}}{4\pi},\qquad f_{h}^{\prime}:=f^{\prime}(r_{h}),\label{eq:local_Hawking}\\
\tilde{M} & \sim & M,\qquad\Phi\sim Q/r_{h}^{d-2}.\label{eq:ADM}
\end{eqnarray}
Here $\Omega_{d-1}^{(k=1,0,-1)}$ is the volume of the ``unit'' sphere, plane
or hyperbola, where for the latter two (noncompact) spaces certain
compactifications are understood.\footnote{{For non-standard values of $k$, the sphere or hyperbola is rescaled, so is its volume $\Omega_{d-1}^{(k)}$. See the discussions in the next section.}} Then the most important question
we are going to answer is: Are there any more charges of the RN-AdS
black hole (\ref{eq:metric}), besides the known ones (the mass $M$
and the electric charge $Q$)?\footnote{The entropy $S$ can be thought of as a Noether charge\citep{Wald}.
One more possible charge is the angular momentum $J$, as will be
mentioned in the Discussion.}

Since the above question eventually leads to an extension of the usual black hole thermodynamics, before trying to answer that question, we should review some other efforts to extend the black hole thermodynamics. The most studied case is to consider the cosmological constant $\Lambda$ as an additional thermodynamic quantity, associated with the so-called thermodynamic volume $\mathcal{V}$ as its conjugate, sometimes called the extended phase space approach (see, e.g. \cite{KRT,Dolan,BFM}). This approach has been further generalized to modified gravitational theories (see e.g. \cite{KRT2} for a generalization to Lovelock gravity). These previous efforts are different from our discussion in that $\mathcal{V}$ (or other quantities conjugate to the Lovelock coupling parameters) is not a (conserved) charge, while we are focusing on all possible charges, as stated above. At the same time, considering $\Lambda$ or other coupling parameters as thermodynamic quantities goes beyond the standard thermodynamics, because they are parameters of the theory itself instead of parameters characterizing different solutions (states) in a fixed theory (system). Although some modifications of the standard Einstein(-Maxwell) gravity are known to treat $\Lambda$ as a parameter of solutions (see, e.g. \cite{Teitelboim}), it is still not clear whether full holographic frameworks exist for such kind of modified theories, which is also one of our main concerns in this paper, because we want to have a holographic interpretation of our thermodynamics (with the proposed new charge) as the real, standard thermodynamics of the dual boundary system.

\section{Topological charge of the black hole from holography}
\label{holography}
One of the most exciting developments in modern theoretical physics
is that the traditional black hole thermodynamics has acquired a new
and interesting interpretation in AdS/CFT\cite{Maldacena,GKP,Witten}. In particular, this interpretation
has been made systematic and universal in much more general settings
of holography\citep{Tian1,Tian2}.

Actually, Strominger and collaborators\citep{Strominger} first established
hydrodynamics on a finite cutoff surface $r=r_{c}$ in the context
of holographic gravity in 2011. It was later realized that this holographic
correspondence on a finite cutoff surface, called a holographic screen,
can be regarded as a natural extension of the AdS/CFT by a very general
bulk/boundary holographic principle\citep{GKP,Witten,Skenderis,Tian1,Tian2}.

Since we will focus on the black hole thermodynamics, here we only
present this general holographic principle at finite temperature.
This principle is just the correspondence between the Euclidean partition
functional of the bulk black hole and the finite temperature generating
functional (for a given ensemble) of the effective field theory on
the holographic screen, namely
\begin{equation}
Z_{FT}[J]=Z_{bulk}[\bar{\phi}(=J)],
\end{equation}
where $J$ is the external source in the field theory, which is identified
with the boundary value $\bar{\phi}$ of some bulk field(s) $\phi$.
Taking logarithms of both sides of the above relation gives
\begin{equation}
Z_{FT}^{c}[J]\approx I_{bulk}^{os}[\bar{\phi}],\label{eq:principle}
\end{equation}
where $Z_{FT}^{c}[J]$ is the connected generating functional and
we have used the semi-classical (or saddle point) approximation of
the quantum gravity to reduce the bulk side down to the on-shell action
of the classical gravity with fixed $\bar{\phi}$. Upon variation
with respect to $J$, the above relation gives the connected Green
function (correlation function) in the field theory in terms of variations
of the bulk on-shell action with respect to $\bar{\phi}$, which is
just the so-called holographic dictionary. In particular, this principle
gives the ordinary AdS/CFT dictionary upon taking into account the
holographic renormalization (see \cite{Skenderis} for a review) to the otherwise divergent on-shell action
for the holographic screen tending to the conformal boundary of an
asymptotic AdS bulk space-time\citep{Tian2}.

For our purpose, we only need to consider the one-point function,
which reads from variation of (\ref{eq:principle}) as
\begin{equation}
\left\langle \mathcal{O}_{\phi}(x)\right\rangle _{FT}=\frac{\delta I_{bulk}^{os}[\bar{\phi}]}{\delta\bar{\phi}(x)},\label{eq:dictionary}
\end{equation}
where the saddle point approximation is always assumed (with ``$\approx$''
suppressed) and 
\[
\mathcal{O}_{\phi}(x)=\frac{\delta I_{FT}[J]}{\delta J(x)}
\]
is the field variable (operator upon quantization) dual to the bulk
field $\phi$. Note that the above expression is just formal, since
in general the action $I_{FT}[J]$ of the boundary field theory cannot
be explicitly written down.

Similar to the standard AdS/CFT, a bulk Maxwell field $A_{\mu}$ is
dual to a global U(1) conserved current $j^{a}$ on the holographic
screen, with the precise correspondence
\begin{equation}
\left\langle j^{a}(x)\right\rangle =\frac{1}{\sqrt{g}}\frac{\delta I_{bulk}^{os}[\bar{A}]}{\delta\bar{A}_{a}(x)}=-n_{\mu}F^{\mu a}(x)|_{{\rm bdry}}
\end{equation}
according to (\ref{eq:dictionary}), where $n^{\mu}$ is the outward
unit normal to the screen. The most important examples (including
the above one) of the dictionary\citep{Tian1,Tian2} that we will
use in the following are listed in Table \ref{tab:dictionary}.
\begin{table}
\caption{\label{tab:dictionary}}

\centering{}%
\begin{tabular}{|c|c|c|}
\hline 
Fields & Bulk & Boundary\tabularnewline
\hline 
\hline 
Maxwell $A_{\mu}$ & $-n_{\mu}F^{\mu a}|_{{\rm bdry}}$ & Current $\left\langle j^{a}\right\rangle $\tabularnewline
\hline 
Gravitational $g_{\mu\nu}$ & Brown-York tensor $t^{ab}|_{{\rm bdry}}$ & Stress tensor $\left\langle t^{ab}\right\rangle $\tabularnewline
\hline 
 & Local Hawking temperature & Temperature\tabularnewline
\hline 
 & Bekenstein-Hawking entropy & Entropy\tabularnewline
\hline 
\end{tabular}
\end{table}

Considering the holographic screen $r=r_{c}$ in the RN-AdS bulk space-time
(\ref{eq:metric}), one can easily show that the Brown-York tensor\citep{BY1}
takes the form of a (relativistic) perfect fluid, from which the energy
density $\epsilon$ and pressure $p$ can be read off as
\begin{equation}\label{energy_pressure}
\epsilon=C-\frac{d-1}{8\pi G}\frac{\sqrt{f_c}}{r_c},\qquad p=\frac{d-2}{8\pi G}\frac{\sqrt{f_c}}{r_c}+\frac{1}{16\pi G}\frac{f_c^\prime}{\sqrt{f_c}}-C,\qquad f_{c}:=f(r_{c}).
\end{equation}
Here $C$ is a constant contributed from considerations of holographic renormalization\cite{Tian2}, which is not important in the following discussion.
Naturally defining the total volume\footnote{Note that the volume $\Omega_{d-1}^{(k)}$ depends on the compactifications
for $k\le0$ as mentioned previously, but hereafter we assume that
this volume is a constant $\Omega_{d-1}=\Omega_{d-1}^{(k=1)}$ under
suitable compactifications {(and rescaling of $k$ if necessary)}\citep{Tian2}.}
\begin{equation}\label{volume}
V=\Omega_{d-1}r_{c}^{d-1}
\end{equation}
and energy
\begin{equation}\label{energy}
E=\epsilon V,
\end{equation}
one obtains the first law of thermodynamics
\begin{equation}
dE+pdV=TdS+\mu dQ,\label{eq:first}
\end{equation}
where the chemical potential
\begin{equation}\label{chemical}
\mu=-\frac{d-1}{8\pi G}\frac{\Omega_{d-1} Q}{\sqrt{f_c}}(\frac{1}{r_c^{d-2}}-\frac{1}{r_h^{d-2}})
\end{equation}
is proportional to the difference of electric potentials between the holographic screen and the horizon\footnote{See the next section for a more precise description.}.
Note that here the temperature $T=T_H/\sqrt{f_c}$ is the local Hawking temperature
on the screen, which differs from the Hawking temperature $T_{H}$
in (\ref{eq:local_Hawking}) by a redshift factor. Actually, a thermodynamic
relation similar to (\ref{eq:first}) has already been obtained in
\citep{BY1,BY2} without referring to holography. Only with the principle
(\ref{eq:principle}) of holographic correspondence does the relation
(\ref{eq:first}) have the meaning of thermodynamics for the dual
field theory on the holographic screen. In more general gravitational
theories (with more general matter content), one can see that this holographic first law (\ref{eq:first})
of thermodynamics still holds by using Hamilton-Jacobi-like analyses\citep{Tian2}.

In the planar case ($k=0$), since the dual field theory lives in
an ordinary flat space-time ($r=r_{c}$), it is expected that the
Gibbs-Duhem relation
\begin{equation}
E+pV=TS+\mu Q\label{eq:Gibbs}
\end{equation}
in the standard thermodynamics should hold, which can be straightforwardly
verified and generally proved by extensibility (or scaling) arguments\citep{Tian2}.
However, it is not the case for the non-planar configuration ($k\ne0$),
which is also unsurprising because of the extra scale introduced by
the spatial curvature of the screen. Actually, from the expressions (\ref{entropy}), (\ref{energy_pressure})-(\ref{energy}), (\ref{chemical}), the condition $f(r_h)=0$ and
\begin{equation}
T=\frac{T_H}{\sqrt{f_c}}=\frac{1}{2\pi\sqrt{f_c}}[\frac{r_h}{\ell^{2}}+(d-2)\frac{M}{r_h^{d-1}}-(d-2)\frac{Q^{2}}{r_h^{2d-3}}],
\end{equation}
after some straightforward arithmetics we have
\begin{equation}\label{Gibbs_check}
E+pV-TS-\mu Q=
\frac{\Omega_{d-1} k}{8\pi G\sqrt{f_c}}(r_{h}^{d-2}-r_c^{d-2}),
\end{equation}
which vanishes only if $k=0$ for $r_c\ne r_h$.

Most interestingly, it turns out that even in the non-planar case,
the Gibbs-Duhem relation (\ref{eq:Gibbs}) can be rectified by introducing
a new ``charge''
\begin{equation}
\varepsilon=\Omega_{d-1}k^{\frac{d-1}{2}}\label{eq:topology}
\end{equation}
and its conjugate quantity
\begin{equation}
\varsigma=(\frac{\partial E}{\partial\varepsilon})_{S,Q,V}=\frac{\varepsilon^{\frac{3-d}{d-1}}\Omega_{d-1}^{\frac{d-3}{d-1}}}{8\pi}\frac{r_{h}^{d-2}-r_{c}^{d-2}}{\sqrt{f_{c}}},\label{eq:potential}
\end{equation}
which means that we have an extended first law
\begin{equation}
dE+pdV=TdS+\mu dQ+\varsigma d\varepsilon.\label{eq:extended}
\end{equation}
From (\ref{Gibbs_check}), (\ref{eq:topology}) and (\ref{eq:potential}), one readily sees that the Gibbs-Duhem-like relation
\begin{equation}
E+pV=TS+\mu Q+\varsigma\varepsilon\label{eq:Gibbs-like}
\end{equation}
still holds. Importantly, it can be shown that for odd $d$ this new
``charge'' $\varepsilon$ is proportional to the Euler number, i.e.
a topological charge, while for even $d$ this interpretation is just
formal\citep{Tian2}.\footnote{Note that if $\varepsilon$ is not defined by (\ref{eq:topology}),
then the relation (\ref{eq:Gibbs-like}) and the first law (\ref{eq:extended})
cannot hold simultaneously, so the validity of the Gibbs-Duhem-like
relation can be viewed as equivalent to the fact that $\varepsilon$
is the topological charge.} Recall that in standard thermodynamics for every conserved charge
we can introduce its conjugate potential, so it is really not strange
that the topological charge $\varepsilon$ and its conjugate quantity
$\varsigma$ play a role in the first law of thermodynamics for
a system in a curved space (with non-trivial topology). In fact, this
new charge $\varepsilon$ just takes into account the extra scale
introduced by the spatial curvature, which makes the usual scaling
arguments valid to produce the Gibbs-Duhem(-like) relation (\ref{eq:Gibbs-like}).\footnote{In other words, the effect of the non-extensibility caused by the spatial curvature can be properly accounted for, at least in the thermodynamics under consideration, by the introduction of the extra charge $\varepsilon$.}
On the other hand, since the scaling arguments use essentially the property of thermodynamic quantities (e.g. the entropy $S$) as homogeneous functions of the complete set of independent extensive variables (basically conserved charges, except the volume $V$, for $S$ being the considered thermodynamic quantity) of the system, the Gibbs-Duhem(-like) relation just fails if we have not taken into account all the conserved charges. Therefore, the relation (\ref{eq:Gibbs-like}) may be thought
of as a completeness relation, which means that we have known all the conserved charges of this holographic black hole system. In this sense, we say that $\varepsilon$ is the last charge of a black hole in the Einstein-Maxwell theory, which we have missed before.

Strictly speaking, the topological charge $\varepsilon$ should be
an integer, as well as the electric charge $Q$ should be quantized,
which makes the first law (\ref{eq:extended}) less interesting. However,
when this law is expressed in terms of densities:
\begin{equation}
d\epsilon=Tds+\mu d\rho+\varsigma de\label{eq:density}
\end{equation}
with the entropy density $s=S/V$, electric charge density $\rho=Q/V$
and topological charge density (Euler density) $e=\varepsilon/V$,
the quantization of $\varepsilon$ and $Q$ can be smoothed out by
a large volume $V$ of the boundary system. {Actually, in the hydrodynamic regime of holography (the fluid/gravity duality)\cite{BHMR}, the system can be inhomogeneous and (\ref{eq:density}) will hold locally. In this case, the Euler density $e$ and its conjugate potential $\varsigma$ are really dynamical quantities.}

\section{Topological charge in the black hole thermodynamics}

The existence and importance of the topological charge of black holes
will be more convincing if it also plays some role in the (ordinary)
black hole thermodynamics. Especially, things will be made more interesting
by considering a slightly generalized framework of the standard black
hole thermodynamics, which enables one to obtain a ``thermodynamic''
relation between off-horizon quantities of the black hole and which
includes the standard black hole thermodynamics as a special (on-horizon)
case.

Basically, there are two approaches to achieve this goal. The first
approach is called ``black hole thermodynamics on equipotential surfaces'',
which is proposed in \citep{CL}. Considering variation between different
configurations of the RN-AdS black hole (\ref{eq:metric}) related
to an equipotential surface $f(r)=c$ with fixed $c$, one easily
obtains the generalized first law
\[
d\tilde{M}=TdS+\Phi dQ
\]
with $\tilde{M}$ the same ADM mass as in (\ref{eq:ADM}),
\begin{equation}
T(r)=\frac{f'(r)}{4\pi}\label{eq:Unruh-Verlinde}
\end{equation}
the Unruh-Verlinde temperature\citep{CL,TW1,TW2},
\begin{equation}
S(r)=\frac{\Omega_{d-1}r^{d-1}}{4}\label{eq:Padmanabhan}
\end{equation}
the Wald-Padmanabhan entropy\footnote{This entropy is the standard Wald entropy generalized to the off-horizon
case by Padmanabhan\citep{Padmanabhan}, which is proportional to
the area of the equipotential surface for the Einstein gravity. See
also \citep{TW1,TW2}.} and
\begin{equation}
\Phi(r)\sim\frac{Q}{r^{d-2}}\label{eq:electric}
\end{equation}
the electric potential at the equipotential surface. When $c=0$,
i.e. $r=r_{h}$, the equipotential surface is just the horizon and
the above first law becomes that of the standard black hole thermodynamics.
This formalism has been generalized to arbitrary dimensions and the
general Lovelock-Maxwell theory in \citep{TW1,TW2}.

Now we illustrate how to incorporate the topological charge into the
above formalism. Writing the gravitational potential $f(r)$ in (\ref{eq:metric})
as
\[
f(r,k,M,Q)=k+\frac{r^{2}}{\ell^{2}}-\frac{2M}{r^{d-2}}+\frac{Q^{2}}{r^{2d-4}},
\]
we have for the equipotential surface $f=c$ (constant)
\[
df(r,k,M,Q)=\frac{\partial f}{\partial r}dr+\frac{\partial f}{\partial k}dk+\frac{\partial f}{\partial M}dM+\frac{\partial f}{\partial Q}dQ=0.
\]
Noting
\begin{eqnarray}
 &  & \partial_{r}f=4\pi T,\qquad\partial_{k}f=1,\nonumber \\
 &  & \partial_{M}f=-\frac{2}{r^{d-2}},\qquad\partial_{Q}f=\frac{2Q}{r^{2d-4}},
\end{eqnarray}
we obtain
\[
dM=2\pi Tr^{d-2}dr+\frac{r^{d-2}}{2}dk+\frac{Q}{r^{d-2}}dQ
\]
Defining the entropy $S$ as in (\ref{eq:Padmanabhan}), we finially
see the generalized first law
\begin{equation}
d\tilde{M}=TdS+\varpi d\varepsilon+\Phi dQ,\qquad\varpi:=\frac{\varepsilon^{\frac{3-d}{d-1}}\Omega_{d-1}^{\frac{-2}{d-1}}}{2\pi(d-1)}S',\label{eq:varpi}
\end{equation}
where $S':=dS/dr$, $\Phi$ is given by (\ref{eq:electric}) and $\varepsilon$
is exactly the topological charge (\ref{eq:topology}). Remarkably,
its appearance in the above first law is not artificial, since the
conjugate quantity (\ref{eq:potential}) in the holographic thermodynamics
is just the difference of the above conjugate potential $\varpi$
(called the topological potential) between the holographic screen
(as the equipotential surface) and the horizon, redshifted by $\sqrt{f_{c}}$.
Actually, as briefly mentioned in the previous section, the structure
for the electric charge is similar\citep{Tian1,Tian2}: The chemical
potential $\mu$ (conjugate to $Q$) in the holographic thermodynamics
is just the difference of the above electric potential $\Phi$ between
the holographic screen and the horizon, redshifted by $\sqrt{f_{c}}$.

The above black hole thermodynamics with the topological charge can
be easily generalized to the Lovelock-Maxwell theory\citep{Tian2},
where the entropy $S$ becomes the Wald-Padmanabhan entropy
\begin{equation}
S=\frac{d-1}{4}\Omega_{d-1}r^{d-1}\sum_{j}\frac{j\tilde{\alpha}_{j}}{d-2j+1}(\frac{k-f}{r^{2}})^{j-1}\label{eq:Lovelock}
\end{equation}
and the topological potential is still given by (\ref{eq:varpi})
but now $S'$ is defined as
\begin{equation}
S':=\frac{d}{dr}S(r,f(r)).\label{eq:S_prime}
\end{equation}
Noting that in the Lovelock-Maxwell case the entropy (\ref{eq:Lovelock})
depends nontrivially on $k$, we again emphasize that the appearance
of $\varepsilon$ in the generalized first law (\ref{eq:varpi}) is
not tuned by hand.

The second approach is called ``black hole thermodynamics from gravitational
equations of motion''. In this approach, we just put the metric of
the maximally symmetric black holes into Einstein's equations (or
more generally the equations of motion of the Lovelock gravity) with
matter, which is then recognized as the first law of thermodynamics
with the topological charge. For simplicity, we take the metric ansatz
\[
ds_{d+1}^{2}=\frac{dr^{2}}{f(r)}-f(r)dt^{2}+r^{2}d\Omega_{d-1}^{(k)2},
\]
which is enough for the Einstein-Maxwell (or Lovelock-Maxwell) case
that we focus on in this essay. The most general case of a maximally
symmetric black hole can also be treated similar to the case without
the topological charge in \citep{TW2}. Upon substitution of the above
ansatz into Einstein's equations
\begin{equation}
R_{\mu\nu}-\frac{1}{2}Rg_{\mu\nu}=8\pi T_{\mu\nu}\label{eq:EOM}
\end{equation}
with $T_{\mu\nu}$ the stress-energy tensor of the matter, the nontrivial
part of them is
\begin{equation}
rf'-(d-2)(k-f)=\frac{16\pi P}{d-1}r^{2}\label{eq:nontrivial}
\end{equation}
with $P=T_{r}^{r}=T_{t}^{t}$ the radial pressure of the matter. Now
we focus on a maximally symmetric screen with fixed $f$ in different
static, maximally symmetric solutions of (\ref{eq:EOM}). In order
to do so, we just need to compare two such configurations of infinitesimal
difference. In fact, multiplying both sides of (\ref{eq:nontrivial})
by the factor
\begin{equation}
\frac{d-1}{16\pi}\Omega_{d-1}r^{d-3}dr,\label{eq:factor}
\end{equation}
we have after some simple algebra (assuming $f$ fixed)
\[
\frac{f'}{4\pi}d\left(\frac{\Omega_{d-1}r^{d-1}}{4}\right)-d\left(\frac{d-1}{16\pi}\Omega_{d-1}(k-f)r^{d-2}\right)+\frac{d-1}{16\pi}\Omega_{d-1}r^{d-2}dk=Pd\left(\frac{\Omega_{d-1}r^{d}}{d}\right).
\]
The above equation is immediately recognized as the first law
\begin{equation}
TdS-dE+\varpi d\varepsilon=Pd\mathcal{V}\label{eq:1st_law}
\end{equation}
with $T$ again the Unruh-Verlinde temperature (\ref{eq:Unruh-Verlinde})
on the screen, $S$ again the Wald-Padmanabhan entropy (\ref{eq:Padmanabhan})
of the screen, $\varpi$ the same topological potential as in (\ref{eq:varpi}),
$\varepsilon$ again the topological charge (\ref{eq:topology}),
$\mathcal{V}=\Omega_{d-1}r^{d}/d$ the volume of the (standard) $d$-``ball'' and
\[
E=\frac{d-1}{16\pi}\Omega_{d-1}(k-f)r^{d-2}
\]
the Misner-Sharp energy inside the screen\citep{Misner-Sharp}.

Then we generalize the above discussion to the Lovelock gravity. In
this case, the nontrivial part of the equations of motion is
\[
\sum_{j}\tilde{\alpha}_{j}(\frac{k-f}{r^{2}})^{j-1}[jrf'-(d-2j)(k-f)]=\frac{16\pi P}{d-1}r^{2}.
\]
Following the same logic as in the Einstein gravity, after multiplying
both sides of the above equation by the same factor (\ref{eq:factor})
we obtain
\begin{eqnarray*}
 &  & \frac{f'}{4\pi}d\left(\frac{d-1}{4}\Omega_{d-1}r^{d-1}\sum_{j}\frac{j\tilde{\alpha}_{j}}{d-2j+1}(\frac{k-f}{r^{2}})^{j-1}\right)\\
 &  & -d\left(\frac{d-1}{16\pi}\Omega_{d-1}r^{d}\sum_{j}\tilde{\alpha}_{j}(\frac{k-f}{r^{2}})^{j}\right)+\frac{d-1}{16\pi}\Omega_{d-1}\sum_{j}j\tilde{\alpha}_{j}(k-f)^{j-1}r^{d-2j}dk\\
 & = & Pd\left(\frac{\Omega_{d-1}r^{d}}{d}\right).
\end{eqnarray*}
The above equation is again recognized as the first law (\ref{eq:1st_law})
with the Wald-Padmanabhan entropy (\ref{eq:Lovelock}) in the Lovelock
gravity and (\ref{eq:S_prime}) in the definition (\ref{eq:varpi})
of the topological potential.

\section{Conclusion and discussion}

To conclude, we have shown in this essay that a maximally symmetric
black hole has a topological charge $\varepsilon$, besides the usual
conserved charges (the mass $M$ and electric charge $Q$) and the
Noether charge $S$. This topological charge has been largely neglected,
if not totally missed, before, but it naturally arises in the holography
with a maximally symmetric screen. In the sense of the Gibbs-Duhem-like
relation (\ref{eq:Gibbs-like}) viewed as a completeness relation,
this topological charge is the last charge of the maximally symmetric
black hole. The existence of this charge and the interesting role
played by it has then been shown in the black hole thermodynamics
from two different approaches, as general as in the Lovelock(-Maxwell)
theory.

In the context of AdS/CFT, actually, there are already discussions about the deviation from the Gibbs-Duhem relation (\ref{eq:Gibbs}) in the case of $k=1$, which is related to the Casimir energy by the Cardy-Verlinde formula (see, for example, \cite{Cai,Gibbons} and references therein). However, it seems not clear the precise relation between those discussions and the topological charge introduced here.

The topological charge with its associated first law of thermodynamics has some direct implications/applications. The first one is to consider how the extended thermodynamics changes the thermodynamic behavior of black holes. Actually, there is already some progress along this direction\cite{Lan}. The second one is to investigate its application in the the fluid/gravity duality, as mentioned at the end of Sec.~\ref{holography}, where the holographic screen can have a curved geometry determined by dynamics, as long as its curvature variation is temporally and spatially slow enough to stay within the hydrodynamic regime.

It is also interesting to ask whether more general black holes, not only
the maximally symmetric ones, have topological charges. Perhaps the
simplest non-maximally symmetric black holes are those with horizons
being direct products of maximally symmetric spaces, but it is not
clear whether holography can be well defined in presence of such bulk
black holes. A more interesting case is a rotating black hole, i.e.
Kerr(-AdS) black hole in Einstein's gravity. In this case, it turns
out that a first-law-like relation including the topological charge
can be written down, but a completeness check similar to the Gibbs-Duhem-like
relation (\ref{eq:Gibbs-like}) is still lacking, again due to the
unclear holography for a black hole with angular momentum. The discussion
of the rotating case is also limited by the fact that up to now we
have no analytical solution of such black holes in the Lovelock(-Maxwell)
theory. Anyway, the topological charge of black holes beyond the maximally
symmetric case is worthy of further investigation.

\begin{acknowledgments}
The author would like to thank Xiaoning Wu, Hongbao Zhang and Liming
Cao for helpful discussions or valuable comments. He also thanks the
Yukawa Institute for Theoretical Physics at Kyoto University, where
this paper was pushed forward during the workshop YITP-T-17-02 ``Gravity
and Cosmology 2018''. This work is partly supported by the National
Natural Science Foundation of China (Grant Nos. 11475179 and 11675015).
It is also supported by the ``Strategic Priority Research Program
of the Chinese Academy of Sciences'', grant No. XDB23030000.
\end{acknowledgments}

\end{document}